Volkan Erol, Okan University Computer Engineering Department, Istanbul, Turkey

volkan.erol@gmail.com

**Entanglement Monotones and Measures: an overview**

**Abstract**

Quantum information theory and quantum computing are theoritical basis of quantum computers. Thanks to entanglement, quantum mechanical systems are provisioned to realize many information processing problems faster than classical counterparts. For example, Shor's factorization algorithm, Grover's search algorithm, quantum Fourrier transformation, etc.. Entanglement, is the theoretical basis providing the expected speedups. It can be view in bipartite or multipartite forms. In order to quantify entanglement, some measures are defined. On the other hand, a general and accepted criterion, which can measure the amount of entanglement of multilateral systems, has not yet been found. In this work, we make a short review of recent research on the topic entanglement monotones/measures with an analitical approach.

## 1- Introduction

Quantum Information is a way to prove the validity of challenging physical experiments. How should computer scientists benefit from the concept of quantum computing until quantum computers are invented? In fact, Quantum Computing alone is not interested in making new Computing Devices. It is becoming a radical Scientific Industry revolutionary journalist to change our point of view of real problems in the world and to find solutions much faster than the present.

However, since it must be used in many information processing tasks, the production and processing of multilateral quantum entangled systems is at the top of the hot topics of recent years [1-8]. Much of the work in the basic quantum technologies, such as quantum cryptography, communications, and computers, requires multi-partite entangled systems such as GHZ, W [9,10]. It can be suggested that the quantum entanglement criteria reflects the different properties of the systems. Many recent research has been done in entanglement and its related disciplines like entanglement measures and majorization, etc. [11-20].

Some swarm-based solutions are included in the current work to speed up machine learning procedures. An example of these studies is entitled "Entanglement-Based Machine Learning on a Quantum" by Cai et al., [21].

Yamomoto and his group have also carried out various studies on the implementation of artificial neural networks in the perspective of Quantum Knowledge Theory [22].

Some of the problems listed above are concepts introduced by the predictions of quantum computers of the last of the contemporary classical cryptography concepts. For example: Quantum Cryptography. The concept of post-quantum cryptograhy has also been introduced along with the opening of the quantum age in the field of cryptography. The basis of this theoretical background is the concept of linear algebraic lattice. According to the current assumptions, a quantitative solution algorithm of lattice-based encryption algorithms has not yet been found. If one of these algorithms can be broken in a quantum way, the post-quantum cryptography concept will be among the dusty shelves of science history.

As it turns out, there are studies using the Quantum Information Theory infrastructure that are closely related to many current fields of Computer Science and Engineering, or it is foreseen that they can be done in a perspective of 10 years. According to the reports of US-based research institutes, which make a respectable trend analysis like Gartner, it is possible to enter the life of a general commercial Quantum Computer in 10 years technological perspective.

## 2- History of Quantum Information Theory

In the past centuries, since the world has been viewed as a deterministic view, real-world problems have been dealt with as if to solve a large clock-like system. Along with the spread of computers in our lives, our understanding of science, mathematics, and collecting has also changed. We are no longer using computers to solve problems, we are building, programming and using computers.

For example, for different types of problems, such as DNA analysis, language processing, and cognitive science, data needs to be transformed to optimize concepts such as compression and error correction of information. It is important to consider concepts such as calculation efficiency, game theory and economic problems. Computer Science has also changed the aims of these and similar fields. In mathematical research, more emphasis is given to efficiency and studies in computer-related fields such as Information Theory, Graph Theory and Statistics have been accelerated. The question of defining *P* or *NP* problems defined between Clay

Millenium problems is trying to explain the oldest puzzle in mathematics: what makes it difficult to find a proof?

When computers came out for the first time, it was hard to imagine that anyone but a few would turn out to be such a big commercial success. This commercial success also led to an intellectual revolution. For example, the invention of the concept of entropy, which is the theoretical basis for data compression or error correction, has been possible with this revolution. This concept was used for the understanding of thermodynamics and steam machines in the 19th century. Claude Shannon is II. He used the concept of entropy in practice during his work on cryptography in Bell Laboratories during World War II. This situation has not only occurred in computer science problems. For example, Einstein used the concept of clock synchronization in his experiments. The problem of clock synchronization was conceived as one of the major industrial problems for that period in terms of automating the movements of trains. In some instances, science has followed technological developments and changed the point of view of the problems after these discoveries.

The story of Quantum Informatics is similar. Quantum mechanics was invented at the beginning of the 20th century and the modern form currently used is known since 1930. However, the idea that quantum mechanics can provide a computational advantage has been put forward much later. This idea emerged when physicists attempted to simulate the quantum mechanics on computers. When they tried it, they faced another problem. A single system (photon polarization) can be described by two complex numbers (the amplitude values of the vertical and horizontal components of polarization), whereas for $n$ systems the number is represented by *2n* rather than $2^n$ complex numbers, and additionally the measurement only reveals $n$ bits. Physicists have developed closed-form solutions to overcome this problem and need a variety of estimation techniques in cases where the number of examined states increases.

The exponential system state space of quantum mechanics has helped them to realize how large and interesting environments nature actually has in terms of computing science. Until then, the concepts of quantum mechanics that were difficult to explain were seen as restrictive items and deficiencies. For example, the Heisenberg Uncertainty Principle was often seen as a restriction on measures. The concept of entanglement as "quantum-based" or philosophy of quantum mechanics has not been studied in detail in terms of operation as much as quantum computation and quantum cryptography concepts were invented in the 1970s and 1980s.

In 1982, Richard Feynman introduced the concept of quantum information, or in other words, the use of quantum mechanical concepts in the field of computational science. The idea is that even if a quantum computer can be invented, it can simulate quantum mechanics much more effectively than conventional computers. This model was formalized by David Deutsch in 1985. It has also been shown for the first time by Deutsch (a computation of two-bit *XOR* values) that a quantum mechanical computer will run faster than a conventional computer. Similar studies have been shown to accelerate over time, for example, by Peter Shor in 1994 when the problem of integers division into multipliers can be done at polynomial time.

In the 1970s, it was suggested by Stephen Wiesner, then a Ph.D. student at that time, that Heisenberg's restrictions on measurement could be used to prevent the learning of confidential messages, but the important scientific journals at that time rejected this work. This issue was first published in 1984 by Charles Bennett and Gilles Brassard as a quantum cryptographic structure. Until 1991, this study was not taken seriously in the scientific mosque until it was realized by them again.

The most important discovery at this point is the formation of the infrastructure of quantum mechanics and quantum cryptography, which are described in the 1950s. It has also been shown that many problems related to the theory of information can be solved much faster by quantum mechanical concepts. Example, Grover's search algorithm, etc. [1]

Today, studies such as Google, Nasa, and many other prestigious universities and research institutes around the world are at full speed. Work on how to physically generate quantum computers has also accelerated in recent years. Historical developments in these areas and explanations of the models used are shared with detailed information in the following sections.

## 3- Definitions About Entanglement Monotones and Measures

In this section, definitions related to the concept of Entanglement are given and after these definitions, explanations about Entanglement Measures are made. Next, the conditions for defining a process as an Entanglement Measure are defined. In the following subsections, detailed explanations were made with the Entanglement Measures, which has been frequently studied in the literature.

## 3.1. Entanglement

One of the starting parts of a study on entanglement measures should be to define the concept of entanglement. It is important to explain how this concept is used. The usefulness of the concept of entanglement is that we say the *Local Operation Classical Communication (LOCC)* constraint, and then we elaborate on a path that we will detail. This constraint makes both our technological and fundamental motivation important because it directly affects the long-distance quantum communication over the systems we examine.

In any quantum communication experiment, we would like to distribute quantum particles among remote laboratories. Perfect entanglement distribution is required for perfect quantum communication [23]. If we can distribute it without a qubit decoherence, we can also perfectly distribute the entanglement it shares. On the contrary, if we can perfectly distribute the circulating system states, we can use teleportation in order to be able to publish quantum system states with fewer classical communications. However, in feasible experiments where we can apply these processes, the noisy effect will prevent us from sending quantum system states over long distances.

To solve this problem, the distribution of the quantum systems must be made over already existing noisy quantum channels; Then it would be appropriate to perform local quantum processes at higher levels in laboratories that are located at distant distances from each other in order to avoid the effects of noise. Because these local quantum operations ('Local Operations-LO') are made in multi-control environments, they are close to the ideal situation and thus the effects of long-distance communication are prevented. It is often not appropriate to run these systems in completely independent environments. In this case, the existing classical communication (CC) can be realized with the existing standard communication technologies. As shown in Figure 1, we can use this communication to coordinate operations in different laboratories.

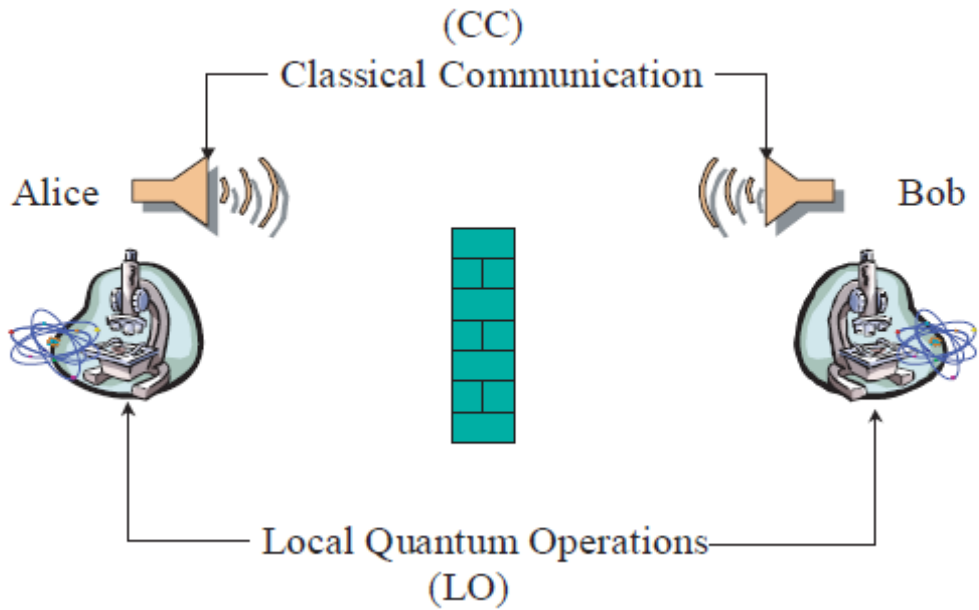

Figure 1- Local Operation (LO) Representation of Classical Communication (CC) method illustration [23]

In many Quantum Information Theory studies it is of vital importance that classical communication can be used, for example quantum teleportation. At present, the assumptions we make are related to the present technological situation, and *LOCC* is a concept that is important in teleportation studies [23].

Entanglement can be defined as quantum correlations between multiple quantum systems. In this case, the question posed is what does quantum correlation look like and what is different from classical correlation? The discussion on 'quantum' and 'classical' effects is a hot topic. We can define classical correlations in the context of quantum information as those arising from the use of *LOCC*. If we look at a quantum system and can not simulate them classically, we generally have *quantum correlations*. Suppose we have a noisy quantum system and we are working on it on LOCC. In this process we can obtain such a system state that we can do some things we can not achieve with classical correlations, such as violating Bell inequality. In this case, we can obtain these effects by quantum correlations in the initial system state that are already present at the source location (even if it is a very noisy system state), not after the LOCC operations. This is the most important point of the entanglement studies.

The limitation of *LOCC* operations is to raise the status of the source system to entangled state. Another definition of entanglement may be that there may not be only a correlation generated by *LOCC* because operations can be performed on non-local binary or multiple quantum systems. In order to be able to understand *LOCC* processes in more detail, Quantum Operations is also described with an entanglement perspective.

## 3.2. Quantum Operations

The studies on quantum information theory generally use 'generalized measurements'. These generalized measures mentioned do not go beyond the standard quantum mechanics. In the general approach to quantum operations, a system changes with respect to a unitary operations, or with projective measurements. We can describe in three steps how a system interacts with other quantum systems in three steps: (1) first we add additional particles (2) then we perform simultaneous unitary and measurement operations on both the system and the particle, and finally (3) we ignore some particles based on the measurement results.

If the additional particles in this process are not originally concerned with the mentioned system, this interaction can be explained by *Kraus operators*. To calculate the total information resulting from any measurement, the measurement result with the probability $\rho_i = tr\{A_i \rho_{in} A_i^\dagger\}$ is calculated as follows:

$$\rho_i = \frac{A_i \rho_{in} A_i^\dagger}{tr\{A_i \rho_{in} A_i^\dagger\}} \qquad (1)$$

Here, $\rho_{in}$ represents the first system state and $A_i$ matrices known as A_i Kraus operators.

The normalization of the probabilities requires that the Kraus operators should have the condition:

$$\sum_i A_i^\dagger A_i = \mathbb{1} \qquad (2)$$

In some cases, for example, if a system interacts with the environment, some or all of the measurement results may not be reached. In the extreme case in this context, the additional particles will trace out.

In this case, the map is given by the following formula: $\sigma = \sum_i A_i \rho_{in} A_i^\dagger$ and is shown in Figure 2 b. This is called the quantum operation which preserves the trace of the map and is often called the *measuring* quantum operation.

Conversely, we can find an operation consisting of the *additional particles*, *equal unitary operation* and *van Neumann measurement* for any $A_i$ linear operation set that yields $\sum_i A_i^\dagger A_i = \mathbb{1}$. For operations that preserve traces, all matrices $A_i$ must be of the same size but $A_i$ may have different sizes if the result information is preserved. Once we have identified the basic building blocks for general quantum operations, we can now define which operations are applicable under *LOCC*. The *LOCC* constraint is visualized in Figure 4.2. In general, this kind of process is very complicated. Alice and Bob can communicate classically before or after a certain number of local movements, in which case any post-lag movements will depend on the results of previous measurements. As a result of this complexity, there is no simple explanation of *LOCC* operations. This motivates the development of easier-to-explain and larger operation classes and remains an integral part of the *LOCC* implementation. One of these important classes is separable operations. Such operations can be written as a product representation in the form of Kraus operations:

$$\rho_k = \frac{A_k \otimes B_k \rho_{in} A_k^\dagger \otimes B_k^\dagger}{tr A_k \otimes B_k \rho_{in} A_k^\dagger \otimes B_k^\dagger} \tag{3}$$

Here $\sum_k A_k^\dagger A_k \otimes B_k^\dagger B_k = \mathbb{1} \otimes \mathbb{1}$ should be satisfied.

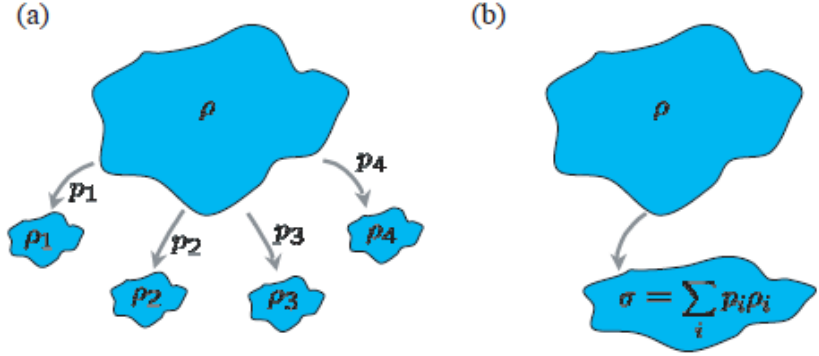

Figure 2-Schematic representation of Quantum Operations in Forms (a) and (b) with sub-selection [23]

Clearly, any *LOCC* operation can be transformed into a separable operation, such that the concatenation operation corresponds to an operation consisting of the multiplications of Alice and Bob's individual local Kraus processes. However, the opposite is not true. A separable operation may not be achieved using LOCC operations.

When separable operations are examined from a mathematical point of view, they can be optimized using separable operations, even though a given task encounters severe limitations using *LOCC*. In some cases, this process can lead to some difficult results: It should be noted that in the case of symmetries, optimally allocable operations can also be achieved with *LOCC*. The general operation classes that support positive partial transpose (PPT) preserving operations create a highly advantageous mathematical model for understanding entanglement.

### 3.3. Definitions for Entanglement

In this section, we define basic definitions about entanglement.

1. Separable states are not entangled:

A state $\rho_{ABC}$ consisting of subparts *A, B, C* and which can be defined as follows are defined as *separable*

$$\rho_{ABC\ldots} = \sum_i p_i \rho_A^i \otimes \rho_B^i \otimes \rho_C^i \otimes \ldots \tag{4}$$

Here $p_i$ is a probability distribution. These system states can easily be created with *LOCC*. Those located in other parts of Alice $p_i$ share the information about the result of users *i* and the user X in each part locally calculates the value of $p_X^i$ and ignores the information that comes at the end of i. Since this system provides states with *LOCC* from the model of local hidden variables, it can be created directly and all correlations of them can be explained classically. In this way, we can arrive at the conclusion that the logically separable system states are not entangled.

2. Non-separable states allow some tasks to be performed better than LOCC, in which case all non-separable system states are entangled:

For any non-separable system state ρ, another such σ system state can be found, which is the result of teleportation fidelity, which can be improved if ρ is present. This interesting result has enabled us to achieve a positive result not in separable system conditions. This also supports the use of non-separable and entanglement terms as synonyms of each other.

3. System states entanglement does not increase under *LOCC* transformations
4. Entanglement does not change under local unitary operations
5. There are some maximally entangled states:

The notion that a system state is entangled allows us to identify in some cases the fact that a system state is more entangled than the other. This leads to the question of the existence of a *maximally entangled state*. The maximally entangled system is more entangled than the others. In this case there could be two-particle two-level systems, or two d-dimensional-level sub-systems called qudit. For a pure system case, the following equation defines the maximally entangled system state:

$$|\psi_d^+\rangle = \frac{|0,0\rangle + |1,1\rangle + \cdots + |d-1,d-1\rangle}{\sqrt{d}} \tag{5}$$

By going all the way out of these definitions; Questions such as *"Is the system state ordering possible?"* Or *"Is the system state ordering problem a partial ordering or is it a complete ordering?"*. In order to be able to query the answers of these questions, one system state needs to be transformed under another *LOCC* procedure and the question must be answered.

## 3.4. Entanglement Measures Postulates

In this section, we will describe a few basic axioms that any entanglement measure should provide. What are the preconditions for a good entanglement measure? A entanglement measure is such a mathematical magnitude that it should provide the fundamental properties associated with the entanglement and should ideally work according to some operational procedures. According to our purposes, this situation allows us to identify some possible desired features. The following list describes the postulates that must be met by the entanglement criteria [45]. Not all of these features are available in many cases:

1- A two-sided (bipartite) entanglement measure $E(\rho)$ is a mapping of positive numbers from system density matrices: $\rho \rightarrow E(\rho) \in \mathbb{R}^+$

    This measure can be defined for the status of any bipartite system. A normalization factor is usually used, for example, for two qudit maximally entangled state

$$|\psi_d^+\rangle = \frac{|0,0\rangle + |1,1\rangle + \cdots + |d-1,d-1\rangle}{\sqrt{d}} \quad (6)$$

    This value is $E(|\psi_d^+\rangle) = \log d$.

2- Is the state is separable than $E(\rho) = 0$.
3- $E$ does not increase in mean under $LOCC$, in other words

$$E(\rho) \geq \sum_i p_i E\left(\frac{A_i \rho A_i^\dagger}{tr A_i \rho A_i^\dagger}\right) \quad (7)$$

    Here $A_i$ represents the *Kraus* operators that define some LOCC protocol, and the probability of i can be calculated with the following equation: $p_i = tr A_i \rho A_i^\dagger$

4- For pure system state $|\psi\rangle\langle\psi|$ measure decreases to entropy of entanglement

$$E(|\psi\rangle\langle\psi|) = (S \circ tr_B)(|\psi\rangle\langle\psi|) \quad (8)$$

Any function E providing first 3 contions are defined as an *entanglement monotone*. Functions conforming conditions 1, 2 and 4 are defined as *entanglement measures*. In litterature, both terms are used as synonymes. There are some additional prerequisites in order to define entanglement measures:

- Convexity

In order for an entanglement measure to provide convexity, it is necessary to provide the following inequality:

$$E(\sum_i p_i \rho_i) \leq \sum_i p_i E(\rho_i) \tag{9}$$

- Additivity

This measure is called additive if $E(\sigma^{\otimes n}) = nE(\sigma)$ equality is provided for every $n$ integers when an entanglement measure and σ system state are defined. This property can not be defined as an essential feature for entanglement measures because there are many entanglement measures that do not provide this property. The more regularized or asymptotic version of this equation can be defined as:

$$E^\infty(\sigma) := \lim_{n \to \infty} \frac{E(\sigma^{\otimes n})}{n} \tag{10}$$

In this case, a measure automatically provides it. As a stronger requirement, we call *full additive* for this measure if we have $E(\sigma \otimes \rho) = E(\sigma) + E(\rho)$ for any $\sigma$ and $\rho$ system state pair.

- Continuity

If an entanglement monotone $L$ is additive for pure states, it conforms the following inequality:

$$n(L(|\phi\rangle)) = L(|\phi\rangle^{\otimes n}) \geq L(\rho_n) \tag{11}$$

This inequality can be written for the entanglement monotones from the third condition. If this equation for $L$ monotone exists

$$L(\rho_n) = L(|\psi^-\rangle^{\otimes nE(\phi))}) + \delta(\epsilon) = nE(\phi)) + \delta(\epsilon) \tag{12}$$

$L$ can be considered "sufficiently continuous". Here $\delta(\varepsilon)$ is a small value and we get the following equation:

$$L(|\phi\rangle) \geq E(|\phi\rangle) + \frac{\delta(\epsilon)}{n} \tag{13}$$

The asymptotic continuous term is defined by the following property:

$$\frac{L(|\phi\rangle_n) - L(|\psi\rangle_n)}{1 + \log(dimH_n)} \to 0 \tag{14}$$

In this case, the two system state flows are the trace norms between $|\phi\rangle_n, |\psi\rangle_n$ $tr||\phi\rangle\langle\phi|_n - ||\varphi\rangle\langle\varphi|_n|$ when $n \to 0$. It is observed that the pure state conditions of L are sufficient and necessary constraints.

## 4. A Short Survey of Commonly Used Entanglement Monotones and Entanglement Measures

In this section, we will describe a number of entanglement measures and monotones defined in the literature for bipartite systems. Some of the measures described here are more physically significant than others. First we will start by defining the concept *of distilable entanglement*:

- Entanglement of Formation:
  The *entanglement of formation* $E_F$ of a mixed state $\rho$, according to Bennett et al. [24,25], is the minimized average entanglement of any ensemble of pure states $|\varphi_i\rangle$ realizing $\rho$:

  $$E_F(\rho) = \inf \sum_i E(|\varphi_i\rangle\langle\varphi_i|) \tag{15}$$

  where infimum is taken over all pure-state decompositions

  $$\rho = \sum_i p_i |\varphi_i\rangle\langle\varphi_i| \tag{16}$$

  and $E(|\varphi_i\rangle\langle\varphi_i|)$ is the entropy of entanglement easily determined by the *von Neumann entropy*. For the special case of two qubits, it is proven by Wootters [26] that the entanglement of formation of a state $\rho$ is given by the formula:

$$E_F(\rho) = H(\frac{1}{2}\left[1 + \sqrt{1 - C^2(\rho)}\right]) \tag{17}$$

where H is the binary entropy, $H(x) = -x\log_2 x - (1-x)\log_2(1-x)$, with the argument related to the Wootters concurence which is defined by [26]:

$$C(\rho) = \max\{0, \lambda_1 - \lambda_2 - \lambda_3 - \lambda_4\} \tag{18}$$

where $\lambda_i$'s are the square roots of the eigenvalues of

$$\rho(\sigma_y \otimes \sigma_y)\rho^*(\sigma_y \otimes \sigma_y) \tag{19}$$

which are in decreasing order and $\sigma_y$ is the Pauli spin matrix and complex conjugation is denoted by *. Both $E_F(\rho)$ and $C(\rho)$ range from *0* for a separable state to *1* for a maximally entangled state.

- Relative Entropy of Entanglement:

*Relative Entropy of Entanglement (REE)* is a measure based on the distance of the state to the closest separable state. Mathematically it can be defined as follows: the minimum of the quantum relative entropy $S(\rho||\sigma) = Tr(\rho\ log\rho - \rho\ log\sigma)$ taken over the set *D* of all separable states $\sigma$, namely for each $\rho$ in *D*

$$E(\rho) = \min_{\rho \epsilon D} S(\rho||\sigma) = S(\rho||\bar{\sigma}) \tag{20}$$

where σ' denotes the closest state to ρ.

For this measure there is no closed formula found for two-level or morelevel systems. For some specific and multi-level systems there are some formula suggestions. For two-level systems there are some estimations based on semidefinite programming [27].

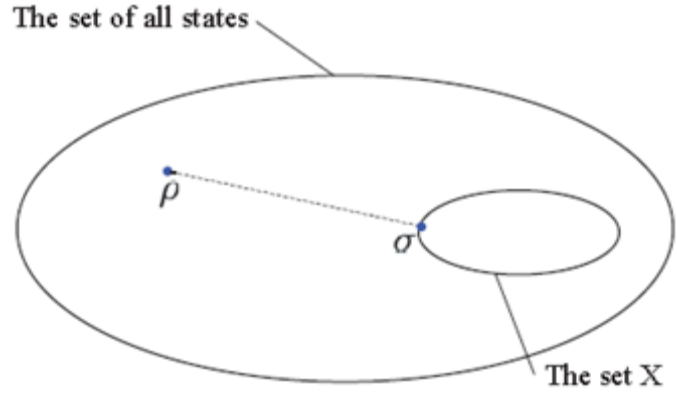

Figure 3: Relative Entropy of Entanglement illustration [23]

- Negativity and Logarithmic Negativity:

Negativity is a quantitative version of Peres-Horodecki criterion [28,29]. It is defined for two particle two level general quantum systems as follows [30-32]:

$$N(\rho) = \max\{0, -2\mu_{min}\} \quad (21)$$

Here $\mu_{min}$ value is the minimum eigenvalue of ρ's partial transpose. Negativity, which is defined by the equation above is a value between 0 and 1 like Concurrence. Similarly like for concurrence, 1 means maximal entanglement. Vidal and Werner shown that Negativity is a monotone function for entanglement [30].

Logarithmic Negativity is calculated with $E_N(\rho) = \log_2(2N(\rho) + 1)$ [33].

Negativity and logarithmic negativity measures are frequently used measures in literature because they are easily calculated measures.

## Conclusion

Studies about Quantum Information Theory continue actively in many research institutions. Algorithms like Shor's factorization algorithm or Grover's search algorithm are shown that should work quite faster on quantum systems compared to classical systems. Very recently, pratical setups of large scale quantum computers are widely studied e.g. quantum repeaters, memories and processors. The doors of a revolunary quantum era in Computer Science is to be opened after some period of time. Technologies like Quantum Key Distribution were defined and developed since many years and they have been daily life products for some sectors like Banking and Military applications.

In Quantum Computing, Entanglement is used for the base computational infrastructure. Entanglement provides us a computational advantage in realization of quantum algorithms. Some ways to quantifiying entanglement were defined. The best formal way to quantify it, is the methods that we call Entanglement Measures or Entanglement Monotones. In this research area, State Ordering Problem is defined and still an open problem especially for multiparticle entangled states. Researchers may deal the definitions of new entanglement measures especially for many-body systems.